\documentclass[aps,prd,groupedaddress,showpacs,nofootinbib]{revtex4}
\usepackage{graphicx,amssymb,amsmath}
\usepackage[english]{babel}

\newcommand{\be}{\begin{equation}}\newcommand{\ee}{\end{equation}}
\newcommand{\bea}{\begin{eqnarray}}\newcommand{\eea}{\end{eqnarray}}
\newcommand{\beaa}{\begin{eqnarray}}\newcommand{\eeaa}{\end{eqnarray}}
\newcommand{\ba}{\begin{array}}\newcommand{\ea}{\end{array}}
\newcommand{\bit}{\begin{itemize}}\newcommand{\eit}{\end{itemize}}
\newcommand{\ben}{\begin{enumerate}}\newcommand{\een}{\end{enumerate}}
\newcommand{\bib}{\bibitem}

\def\lan{\langle}
\def\lf{\left}

\def\non{\nonumber}
\def\ran{\rangle}

\def\ri{\right}

\def\al{\alpha}\def\bt{\beta}
\def\de{\delta}
\def\te{\theta}

\def\si{\sigma}
\def\om{\omega}

\def\1{{_{1}}}\def\2{{_{2}}}

\begin{document}

\title{Dark energy, cosmological constant and neutrino mixing}

\author{A.Capolupo${}^{\natural}$, S.Capozziello${}^{\sharp}$, G.Vitiello${}^{\flat}$}


\affiliation{${}^{\natural}$ Department of Physics and Astronomy,
University of Leeds, Leeds LS2 9JT UK,
\\  ${}^{\flat}$
Dipartimento di Matematica e Informatica,
 Universit\`a di Salerno and Istituto Nazionale di Fisica Nucleare,
 Gruppo Collegato di Salerno, 84100 Salerno, Italy,
\\ ${}^{\sharp}$ Dipartimento di Scienze Fisiche, Universit\`a di Napoli "Federico II" and INFN Sez. di Napoli,
Compl. Univ. Monte S. Angelo, Ed.N, Via Cinthia, I-80126 Napoli,
Italy.}


\vspace{2mm}

\begin{abstract}

The today estimated value of dark energy can be
achieved by the vacuum condensate induced by neutrino mixing
phenomenon. Such a tiny value is recovered for a cut-off of the
order of Planck scale and it is linked to the sub-eV neutrino mass scale.
Contributions to dark energy from auxiliary fields or mechanisms are not necessary in
this approach.

\end{abstract}


\maketitle

\section{Introduction}

The neutrino mixing phenomenon, was firstly studied in the context of quantum mechanics
\cite{Pontecorvo:1957cp,Maki:1962mu,Fujii64,Gribov:1968kq,Bilenky:1978nj,Bilenky,Mohapatra:1991ng,Wolfenstein:1977ue,Giunti:1991ca},
and more recently analyzed in the framework of the quantum field theory (QFT) formalism
\cite{BV95,BHV98,Fujii:1999xa,JM01,JM011,hannabuss,yBCV02,BCRV01,Capolupo:2004pt,Capolupo:2004av,Blasone:2005ae,Blasone:2006jx}.

The recent experimental achievements proving neutrino oscillations
\cite{SNO,K2K} and the progresses in the QFT  understanding
\cite{Blasone:2005ae,Blasone:2006jx} of the neutrino mixing seems to
indicate a promising path beyond the Standard Model of electro-weak
interaction for elementary particles and a possible link between
high energy physics and cosmology
\cite{Blasone:2004yh,Capolupo:2006et}. In this paper, indeed, we
show that the energy content of the neutrino mixing vacuum
condensate \cite{Blasone:2004yh} can be interpreted as dynamically
evolving dark energy \cite{Capolupo:2006et} that, at present epoch,
assumes the behavior and the value of the observed cosmological
constant. We compute such a value and show that, above a threshold,
it is slowly diverging and its derivative with respect to the
cut-off value goes actually to zero (cfr. Fig.(2)), which allows to
use the cut-off at its Planck scale value.

Our result links together dark energy with the sub-eV neutrino mass
scale. The link comes from the neutrino-antineutrino pair vacuum
condensate due to the mixing phenomenon.

This fact is crucial from a genuine experimental point of view
since, up to now, none of the exotic candidates for dark matter and
dark energy, has been detected at a fundamental level. Considering
neutrino mixing vacuum condensate as the source of dark energy fits
with the conservative view by which only actually observed
ingredients as gravity, radiation, neutrinos and baryons are taken
into account.

The layout of the paper is the following. In Section II, we outline
the neutrino mixing formalism in Quantum Field Theory. In Section
III we compute the neutrino mixing contribution to the dark energy
in the case of two generations. The case of the three flavor fermion
mixing is analyzed in Section IV and conclusions are drawn in
Section V.

\section{Neutrino mixing in Quantum Field Theory}

For the reader convenience the main features of the QFT formalism
for the neutrino mixing are here summarized. For the sake of
simplicity, we restrict to the case of two flavors. Extension to
three flavors \cite{yBCV02} can be also considered (for a detailed
review see \cite{Capolupo:2004av}).

The Pontecorvo mixing transformations for two Dirac neutrino fields are
\begin{eqnarray} \nonumber\label{mix}
\nu_{e}(x) &=&\nu_{1}(x)\,\cos\theta + \nu_{2}(x)\,\sin\theta
\\
\nu_{\mu}(x) &=&-\nu_{1}(x)\,\sin\theta + \nu_{2}(x)\,\cos\theta
\;,\end{eqnarray}
where $\nu_{e}(x)$ and $\nu_{\mu}(x)$ are the
fields with definite flavors, $\theta$ is the mixing angle and
$\nu_1$ and $\nu_2$ are the fields with definite masses $m_{1} \neq m_{2}$:
\bea\label{freefi}
 \nu _{i}(x)=\frac{1}{\sqrt{V}}{\sum_{{\bf k} ,
r}} \left[ u^{r}_{{\bf k},i}\, \al^{r}_{{\bf k},i}(t) +
v^{r}_{-{\bf k},i}\, \bt^{r\dag}_{-{\bf k},i}(t) \ri] e^{i {\bf
k}\cdot{\bf x}},\qquad \, \qquad i=1,2,
\eea
with  $ \al_{{\bf k},i}^{r}(t)=\al_{{\bf
k},i}^{r}\, e^{-i\omega _{k,i}t}$,
$ \bt_{{\bf k},i}^{r\dag}(t) = \bt_{{\bf k},i}^{r\dag}\, e^{i\omega_{k,i}t},$
and $ \omega _{k,i}=\sqrt{{\bf k}^{2} + m_{i}^{2}}.$
The operators $\alpha ^{r}_{{\bf k},i}$ and $ \beta ^{r }_{{\bf k},i}$, $
i=1,2 \;, \;r=1,2$ annihilate the vacuum
state $|0\rangle_{1,2}\equiv|0\rangle_{1}\otimes |0\rangle_{2}$:
$\alpha ^{r}_{{\bf k},i}|0\rangle_{12}= \beta ^{r }_{{\bf
k},i}|0\rangle_{12}=0$.
 The anticommutation relations are:
$\left\{ \nu _{i}^{\alpha }(x),\nu _{j}^{\beta
\dagger }(y)\right\} _{t=t^{\prime }}=\delta ^{3}({\bf x-y})\delta
_{\alpha \beta } \delta _{ij},$ with $\alpha ,\beta =1,...4,$ and $\left\{ \alpha _{{\bf k},i}^{r},\alpha _{{\bf
q},j}^{s\dagger }\right\} =\delta _{{\bf kq}}\delta _{rs}\delta
_{ij};$ $\left\{ \beta _{{\bf k},i}^{r},\beta _{{\bf
q,}j}^{s\dagger }\right\} =\delta _{{\bf kq}}\delta _{rs}\delta
_{ij},$ with $i,j=1,2.$
All other anticommutators are zero. The orthonormality and
completeness relations are:
$u_{{\bf k},i}^{r\dagger }u_{{\bf k},i}^{s} = v_{{\bf
k},i}^{r\dagger }v_{{\bf k},i}^{s} = \delta _{rs},\; $
$u_{{\bf k},i}^{r\dagger }v_{-{\bf k},i}^{s} = v_{-{\bf k}
,i}^{r\dagger }u_{{\bf k},i}^{s} = 0,\;$
and
$\sum_{r}(u_{{\bf k},i}^{r}u_{{\bf k},i}^{r\dagger
}+v_{-{\bf k},i}^{r}v_{-{\bf k},i}^{r\dagger }) = 1.$

The mixing transformation
Eqs.(\ref{mix}) can be written as \cite{BV95}:
\bea \label{mixG} \nu_{e}^{\alpha}(x) = G^{-1}_{\bf \te}(t)\;
\nu_{1}^{\alpha}(x)\; G_{\bf \te}(t) \\ \non \nu_{\mu}^{\alpha}(x)
= G^{-1}_{\bf \te}(t)\; \nu_{2}^{\alpha}(x)\; G_{\bf \te}(t) \eea
where the mixing generator $G_{\bf \te}(t)$ is given by
\bea\label{generator12} G_{\bf \te}(t) = exp\left[\theta \int
d^{3}{\bf x} \left(\nu_{1}^{\dag}(x) \nu_{2}(x) - \nu_{2}^{\dag}(x)
\nu_{1}(x) \right)\right]\;.
\eea
At finite volume, $G_{\bf \te}(t)$ is an unitary operator,
$G^{-1}_{\bf \te}(t)=G_{\bf -\te}(t)=G^{\dag}_{\bf \te}(t)$,
preserving the canonical anticommutation relations; $G^{-1}_{\bf
\te}(t)$ maps the Hilbert spaces for free fields ${\cal H}_{1,2}$ to
the Hilbert spaces for interacting fields ${\cal H}_{e,\mu}$: $
G^{-1}_{\bf \te}(t): {\cal H}_{1,2} \mapsto {\cal H}_{e,\mu}.$ In
particular, for the vacuum $|0 \rangle_{1,2}$ we have, at finite
volume $V$:
\bea\label{flavvac}
 |0(t) \rangle_{e,\mu} = G^{-1}_{\bf \te}(t)\;
|0 \rangle_{1,2}\;.
\eea
$|0 \rangle_{e,\mu}$ is the vacuum for ${\cal H}_{e,\mu}$, which
we will refer to as the flavor vacuum.
In the infinite volume limit the flavor vacuum $|0(t) \rangle_{e,\mu}$
turns out to be unitary inequivalent to the vacuum for the massive neutrinos
$|0 \rangle_{1,2}$ \cite{BV95}.
 This can be proved for any number of generations \cite{hannabuss}.
 The non-perturbative nature of the flavored vacuum for the mixed neutrinos is
 thus revealed.

 Due to the linearity of
$G_{\bf \te}(t)$, we can define the flavor annihilators, relative
to the fields $\nu_{e}(x)$ and $\nu_{\mu}(x)$ at each time
expressed as (we use $(\sigma,i)=(e,1) , (\mu,2)$):
\begin{eqnarray}\label{flavannich}
\alpha _{{\bf k},\sigma}^{r}(t) &\equiv &G^{-1}_{\bf \te}(t)\;\alpha
_{{\bf k},i}^{r}(t)\;G_{\bf \te}(t),  \nonumber
\\
\beta _{{\bf k},\sigma}^{r}(t) &\equiv &G^{-1}_{\bf \te}(t)\;\beta
_{{\bf
k},i}^{r}(t)\;G_{\bf \te}(t).
\end{eqnarray}

The flavor fields can be expanded in the same bases as $\nu_{i}$:
\begin{eqnarray}
\nu _{\sigma}({\bf x},t) &=&\frac{1}{\sqrt{V}}{\sum_{{\bf k},r} }
e^{i{\bf k.x}}\left[ u_{{\bf k},i}^{r} \alpha _{{\bf
k},\sigma}^{r}(t) + v_{-{\bf k},i}^{r} \beta _{-{\bf k},\sigma}^{r\dagger
}(t)\right].
\end{eqnarray}

The flavor annihilation operators
in the reference frame
such that ${\bf k}=(0,0,|{\bf k}|)$ are:
\bea\label{annihilator} \non
\alpha^{r}_{{\bf
k},e}(t)&=&\cos\theta\;\alpha^{r}_{{\bf
k},1}(t)\;+\;\sin\theta\;\left( |U_{{\bf k}}|\; \alpha^{r}_{{\bf
k},2}(t)\;+\;\epsilon^{r}\; |V_{{\bf k}}|\; \beta^{r\dag}_{-{\bf
k},2}(t)\right)
\\ \non
\alpha^{r}_{{\bf k},\mu}(t)&=&\cos\theta\;\alpha^{r}_{{\bf
k},2}(t)\;-\;\sin\theta\;\left( |U_{{\bf k}}|\; \alpha^{r}_{{\bf
k},1}(t)\;-\;\epsilon^{r}\; |V_{{\bf k}}|\; \beta^{r\dag}_{-{\bf
k},1}(t)\right)
\\
\beta^{r}_{-{\bf k},e}(t)&=&\cos\theta\;\beta^{r}_{-{\bf
k},1}(t)\;+\;\sin\theta\;\left( |U_{{\bf k}}|\; \beta^{r}_{-{\bf
k},2}(t)\;-\;\epsilon^{r}\; |V_{{\bf k}}|\; \alpha^{r\dag}_{{\bf
k},2}(t)\right)
\\ \non
\beta^{r}_{-{\bf k},\mu}(t)&=&\cos\theta\;\beta^{r}_{-{\bf
k},2}(t)\;-\;\sin\theta\;\left( |U_{{\bf k}}|\; \beta^{r}_{-{\bf
k},1}(t)\;+\;\epsilon^{r}\; |V_{{\bf k}}|\; \alpha^{r\dag}_{{\bf
k},1}(t)\right),
\eea
with $\epsilon^{r}=(-1)^{r}$ and
\bea\label{Vk2}
 |U_{{\bf k}}| \equiv  u^{r\dag}_{{\bf
k},i} u^{r}_{{\bf k},j} = v^{r\dag}_{-{\bf
k},i} v^{r}_{-{\bf k},j}\,,
\qquad \qquad
|V_{{\bf k}}| \equiv  \epsilon^{r}\; u^{r\dag}_{{\bf
k},1} v^{r}_{-{\bf k},2} = -\epsilon^{r}\; u^{r\dag}_{{\bf
k},2} v^{r}_{-{\bf k},1}
\eea
with $i,j = 1,2$ and $ i \neq j$. We have:

  \bea
\non |U_{{\bf
k}}|=\left(\frac{\omega_{k,1}+m_{1}}{2\omega_{k,1}}\right)^{\frac{1}{2}}
\left(\frac{\omega_{k,2}+m_{2}}{2\omega_{k,2}}\right)^{\frac{1}{2}}
\left(1+\frac{{\bf
k}^{2}}{(\omega_{k,1}+m_{1})(\omega_{k,2}+m_{2})}\right)
\\
\label{Vk}|V_{{\bf
k}}|=\left(\frac{\omega_{k,1}+m_{1}}{2\omega_{k,1}}\right)^{\frac{1}{2}}
\left(\frac{\omega_{k,2}+m_{2}}{2\omega_{k,2}}\right)^{\frac{1}{2}}
\left(\frac{k}{(\omega_{k,2}+m_{2})}-\frac{k}{(\omega_{k,1}+m_{1})}\right)
\eea
\bea
|U_{{\bf k}}|^{2}+|V_{{\bf k}}|^{2}=1.
\eea

The condensation density is given by
\bea _{e,\mu}\langle 0| \al_{{\bf k},i}^{r \dag} \al^r_{{\bf k},i}
|0\rangle_{e,\mu}\,= \;_{e,\mu}\langle 0| \bt_{{\bf k},i}^{r \dag}
\bt^r_{{\bf k},i} |0\rangle_{e,\mu}\,=\, \sin^{2}\te\; |V_{{\bf
k}}|^{2} \;, \qquad i=1,2\,. \eea

The Bogoliubov coefficient $|V_{{\bf k}}|^{2}$ appearing in the condensation density
can be written as a function of the dimensionless momentum
$p=\frac{|{\bf k}|}{\sqrt{m_1 m_2}}$ and dimensionless
parameter $a= \frac{(m_{2}-m_{1})^2}{m_1 m_2}$, as follows,
\bea
|V(p,a)|^2 & =& \frac{1}{2}\lf(1-\frac{p^2+1} {\sqrt{(p^2 + 1)^2
+ a p^2}}\ri) ~.\label{Vpa}
 \eea
From Fig.(1) we see that the effect is maximal when $p=1$, and
$|V|^2$ goes to zero for large momenta (i.e. for $|{\bf k}|^2\gg
m_{1} m_{2}$ ) as $|V|^2 \approx \frac{(\Delta m)^2}{4 k^2}$.

\begin{figure}
\centering \resizebox{8.5cm}{!}{\includegraphics{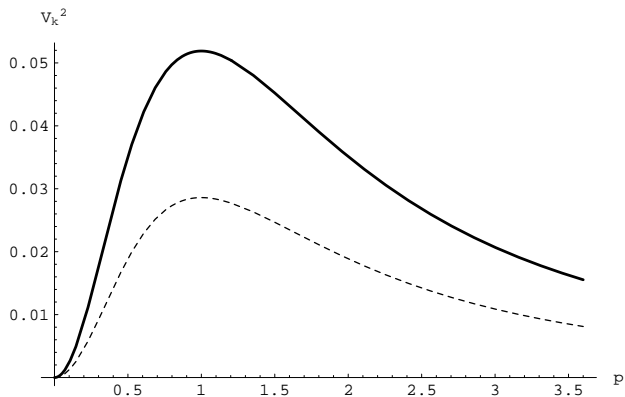}}
\caption{The fermion condensation density $|V(p,a)|^2$ as a
function of $p$ for $a=0.98$ (solid line) and $a=0.5$ (dashed
line).} \label{Fig: 1}
\end{figure}

Since the experimentally observed neutrinos
are always extremely relativistic, the value of $|V|^2$ is very small. Only
for extremely low energies (like those in
neutrino cosmological background) $|V|^2$
might be large and account for few percent.

In the next Section we will show that the mixing of neutrinos may
contribute to the value of the dark energy exactly because of the
non-zero value of $|V_{\bf k}|^2$: its behavior at very high
momenta, together with the Lorentz invariance of the vacuum
condensate at the present time, can be responsible of the very
tiny value of the cosmological constant.

\section{Neutrino mixing and dark energy}

Experimental data indicate that the today observed universe can be
described as an accelerating Hubble fluid where the
contribution of dark energy component to the total matter-energy
density is $\Omega_{\Lambda}\simeq 0.7$ (see the discussion in Section V).
Moreover, the cosmic flow is "today"
accelerating while it was not so at intermediate redshift $z$
(e.g. $1 < z < 10$) where large scale structures have supposed to
be clustered. Thus, physically motivated cosmological models
should  undergo, at least,  three phases: an early accelerated
inflationary phase, an intermediate standard matter dominated
(decelerated) phase  and a final, today observed, dark energy
dominated  (accelerated) phase.
 This means that we have to take into account some
form of {\it dark energy} which evolves from early epochs inducing
the today observed acceleration.

In this Section we  show that the energy density due to neutrino mixing vacuum condensate
can be interpreted as an evolving dark energy that at present epoch has
a behavior and a value compatible with the observed cosmological constant.

The calculation here presented is performed in a Minkowski space-time but it can
be easily extended to curved space-times. When particle mixing and
oscillations in curved background is analyzed, neutrino mixing,
and in general particle mixing, gives a time dependent dark energy
leading, however, to the same final result: the today observed cosmological
dark energy value can be recovered.

Let us calculate the contribution $\rho_{vac}^{mix}$ of the
neutrino mixing to the vacuum energy density.

As well known \cite{Itz}, the Lorentz invariance of the vacuum
implies that the vacuum energy-momentum tensor is equal to zero:
${\cal T}_{\mu\nu}^{vac} = \lan 0 |:{\cal T}_{\mu\nu}:| 0\ran= 0$,
(as usual normal ordering is denoted by the colon $:...:$). The
(0,0) component of the energy-momentum tensor density
 ${\cal T}_{00}(x) $ for the fields $\nu_1$ and $\nu_2$ is then
\bea\
 :{\cal T}_{00}(x): = \frac{i}{2}:\left({\bar \Psi}_{m}(x)\gamma_{0}
\stackrel{\leftrightarrow}{\partial}_{0} \Psi_{m}(x)\right): \eea
where  $\Psi_{m} = (\nu_1, \nu_2)$.
In terms of the annihilation and creation operators of fields
$\nu_{1}$ and $\nu_{2}$, the (0,0) component of the
energy-momentum tensor $ T_{00}=\int d^{3}x {\cal T}_{00}(x)$ is
given by
\bea\label{T00}
 :T^{00}_{(i)}:= \sum_{r}\int d^{3}{\bf k}\,
\omega_{k,i}\lf(\al_{{\bf k},i}^{r\dag} \al_{{\bf k},i}^{r}+
\beta_{{\bf -k},i}^{r\dag}\beta_{{\bf -k},i}^{r}\ri), \eea with
$i=1,2$. Note that $T^{00}_{(i)}$ is time independent.

In the early universe epochs, when the Lorentz invariance of the
vacuum condensate is broken,
 $\rho_{vac}^{mix}$ presents also space-time dependent condensate
 contributions. This  implies that the contribution $\rho_{vac}^{mix}$ of the
neutrino mixing to the vacuum energy density is given by computing
the expectation value of $T^{00}_{(i)}$ in the flavor vacuum $|0(t)
{\rangle}_{e,\mu}$:
 \bea\
\rho_{vac}^{mix} = \frac{1}{V}\; \eta_{00}\; {}_{e,\mu}\lan 0(t)
|\sum_{i} :T^{00}_{(i)}(0):| 0(t)\ran_{e,\mu}   ~.
 \eea

Within the QFT formalism for neutrino mixing, we have
 \bea
 {}_{e,\mu}\lan 0 |:T^{00}_{(i)}:| 0\ran_{e,\mu}={}_{e,\mu}\lan
0(t) |:T^{00}_{(i)}:| 0(t)\ran_{e,\mu}
 \eea
for any t. We then obtain
\bea\non \rho_{vac}^{mix} &= &\sum_{i,r}\int \frac{d^{3}{\bf
k}}{(2 \pi)^{3}} \, \omega_{k,i}\Big({}_{e,\mu}\lan 0 |\al_{{\bf
k},i}^{r\dag} \al_{{\bf k},i}^{r}| 0\ran_{e,\mu} + {}_{e,\mu}\lan
0 |\beta_{{\bf k},i}^{r\dag} \beta_{{\bf k},i}^{r}| 0\ran_{e,\mu}
\Big) \,, \eea
and then

\bea\label{aspT}
 \rho_{vac}^{mix}
&=&\,4\sin^{2}\theta \int \frac{d^{3}{\bf k}}{(2 \pi)^{3}}
\lf(\omega_{k,1}+\omega_{k,2}\ri) |V_{\bf k}|^{2} , \eea
which, introducing the cut-off $K$, becomes
\bea\label{cc}
 \rho_{vac}^{mix} = \frac{ 2}{\pi} \sin^{2}\theta
\int_{0}^{K} dk \, k^{2}(\omega_{k,1}+\omega_{k,2}) |V_{\bf
k}|^{2} \,. \eea

In a similar way, the contribution
 $ p_{vac}^{mix}$ of the neutrino mixing
to the vacuum pressure is given by the expectation value of
$T^{jj}_{(i)}$ (where no summation on the index $j$ is intended)
on the flavor vacuum $| 0\ran_{e,\mu}$:
 \bea\
p_{vac}^{mix}= -\frac{1}{V}\; \eta_{jj} \; {}_{e,\mu}\lan 0
|\sum_{i} :T^{jj}_{(i)}:| 0\ran_{e,\mu} ~.
 \eea
Being
 \bea\label{Tjj}
 :T^{jj}_{(i)}:= \sum_{r}\int d^{3}{\bf k}\, \frac{k^j
k^j}{\;\omega_{k,i}}\lf(\al_{{\bf k},i}^{r\dag} \al_{{\bf
k},i}^{r}+ \beta_{{\bf -k},i}^{r\dag}\beta_{{\bf -k},i}^{r}\ri),
\eea
 in the case of the
isotropy of the momenta we have $T^{11} = T^{22} = T^{33}$, then
 \bea\label{cc2}
  p_{vac}^{mix} = \frac{2}{3\;\pi}
\sin^{2}\theta \int_{0}^{K} dk \,
k^{4}\lf[\frac{1}{\omega_{k,1}}+\frac{1}{\omega_{k,2}}\ri] |V_{\bf
k}|^{2}\,.
 \eea

From Eqs.(\ref{cc}) and (\ref{cc2}) we have that the adiabatic index is $w = p_{vac}^{mix}/
\rho_{vac}^{mix} \simeq 1/3$ when the cut-off is chosen to be $K
\gg m_{1}, m_{2} $.

The values of $\rho_{vac}^{mix}$ and  $ p_{vac}^{mix}$ which we
obtain are time-independent since we are taking into account the
Minkowski metric. Considering a curved space-time,
time-dependence has to be taken into account but the essence of
the result is the same.
 At the present epoch, the
breaking of the Lorentz invariance is negligible and then
 $\rho_{vac}^{mix}$ comes from space-time independent condensate contributions
 (i.e. the contributions carrying a non-vanishing $\partial_{\mu} \sim k_{\mu}=(\omega_{k},k_{j}) $
are missing).
 That is, the energy-momentum density tensor of the vacuum condensate
is given by
\bea {}_{e,\mu}\lan 0 |:T_{\mu\nu}:|
0\ran_{e,\mu} = \eta_{\mu\nu}\;\sum_{i}m_{i}\int
\frac{d^{3}x}{(2\pi)^3}\;{}_{e,\mu}\lan 0 |:\bar{\nu
}_{i}(x)\nu_{i}(x):| 0\ran_{e,\mu}\, =
\eta_{\mu\nu}\;\rho_{\Lambda}^{mix}.
 \eea

Since $\eta_{\mu\nu} = diag (1,-1,-1,-1)$ and, in a homogeneous and
isotropic universe, the energy-momentum tensor is $T_{\mu\nu} = diag
(\rho\,,p\,,p\,,p\,)$, then, consistently with Lorentz invariance,
the state equation is $\rho_{\Lambda}^{mix} = -p_{\Lambda}^{mix}$.
This means that  the vacuum condensate, coming from  neutrino
mixing, contributes today to the dynamics of the universe by a
 cosmological constant behavior {\cite{Capolupo:2006et}. Explicitly, we
have
\bea\label{cost}
\rho_{\Lambda}^{mix}= \frac{2}{\pi}
\sin^{2}\theta \int_{0}^{K} dk \,
k^{2}\lf[\frac{m_{1}^{2}}{\omega_{k,1}}+\frac{m_{2}^{2}}
{\omega_{k,2}}\ri] |V_{\bf k}|^{2}.
\eea
Solving the integral, we
obtain
\bea
\non \rho_{\Lambda}^{mix} &=& \frac{2}{\pi}
\sin^{2}\theta \Big\{(m_{2}^{2}-m_{1}^{2}) k
\Big(\sqrt{k^{2}+m_{2}^{2}}- \sqrt{k^{2}+m_{1}^{2}}\Big)+
\frac{2(m_{2}-m_{1})}{\sqrt{m_{2}^{2}-m_{1}^{2}}}
\Big[m_{1}^{4}\arctan
\Big(\frac{\sqrt{m_{2}^{2}-m_{1}^{2}}}{m_{1}\sqrt{k^{2}+m_{2}^{2}}}k\Big)
\\\non
&-& m_{2}^{4}\arctan \Big(\frac{\sqrt{m_{2}^{2}-m_{1}^{2}}}{m_{2}\sqrt{k^{2}+m_{1}^{2}}}k\Big) \Big]
+ (2 m_{1}^{4}- 2 m_{1}^{3}m_{2}+m_{1}^{2}m_{2}^{2}- m_{2}^{4})\log\lf(k+\sqrt{k^{2}+m_{2}^{2}}\ri)
\\
&+& (2 m_{2}^{4}- 2 m_{2}^{3}m_{1}+m_{1}^{2}m_{2}^{2}- m_{1}^{4})\log\lf(k+\sqrt{k^{2}+m_{1}^{2}}\ri) \Big\}_{0}^{K},
\eea
that is
\bea\non\label{dark}
\rho_{\Lambda}^{mix} &=& \frac{2}{\pi} \sin^{2}\theta
\Big\{(m_{2}^{2}-m_{1}^{2}) K \Big(\sqrt{K^{2}+m_{2}^{2}}-
\sqrt{K^{2}+m_{1}^{2}}\Big)+ \frac{2(m_{2}-m_{1})}{\sqrt{m_{2}^{2}-m_{1}^{2}}}
\Big[m_{1}^{4}\arctan \Big(\frac{\sqrt{m_{2}^{2}-m_{1}^{2}}}{m_{1}\sqrt{K^{2}+m_{2}^{2}}}K\Big)
\\\non
&-& m_{2}^{4}\arctan \Big(\frac{\sqrt{m_{2}^{2}-m_{1}^{2}}}{m_{2}\sqrt{K^{2}+m_{1}^{2}}}K\Big) \Big]
+ (2 m_{1}^{4}- 2 m_{1}^{3}m_{2}+m_{1}^{2}m_{2}^{2}- m_{2}^{4})\log\lf(K+\sqrt{K^{2}+m_{2}^{2}}\ri)
\\\non
&+& (2 m_{2}^{4}- 2 m_{2}^{3}m_{1}+m_{1}^{2}m_{2}^{2}- m_{1}^{4})\log\lf(K+\sqrt{K^{2}+m_{1}^{2}}\ri)
- (2 m_{1}^{4}- 2 m_{1}^{3}m_{2}+m_{1}^{2}m_{2}^{2}- m_{2}^{4})\log\lf(m_{2}\ri)
\\
&-&(2 m_{2}^{4}- 2 m_{2}^{3}m_{1}+m_{1}^{2}m_{2}^{2}- m_{1}^{4})\log\lf(m_{1}\ri)\Big\}.
\eea

The plot of $\rho_{\Lambda}^{mix}$ as function of the momentum
cut-off $K$ (Fig.2) shows that for $K$ at Planck scale a value of
$\rho_{\Lambda}^{mix}$ is obtained, which is in agreement with the
observed value of cosmological constant.

To better understand the meaning of Eq.(\ref{dark}), we report the behavior of
$\rho_{\Lambda}^{mix}$ for $K \gg m_{1},m_{2} $:
\bea\non\label{darklimit}
\rho_{\Lambda}^{mix} & \approx & \frac{2}{\pi} \sin^{2}\theta
\Big\{\frac{2(m_{2}-m_{1})}{\sqrt{m_{2}^{2}-m_{1}^{2}}}
\Big[m_{1}^{4}\arctan \Big(\frac{\sqrt{m_{2}^{2}-m_{1}^{2}}}{m_{1}}\Big)
- m_{2}^{4}\arctan \Big(\frac{\sqrt{m_{2}^{2}-m_{1}^{2}}}{m_{2}}\Big) \Big]
\\
\non &-& (2 m_{1}^{4}- 2 m_{1}^{3}m_{2}+m_{1}^{2}m_{2}^{2}- m_{2}^{4})\log\lf(m_{2}\ri)
- (2 m_{2}^{4}- 2 m_{2}^{3}m_{1}+m_{1}^{2}m_{2}^{2}- m_{1}^{4})\log\lf(m_{1}\ri)
\\ &+& (m_{1}^{4}+m_{2}^{4}+2m_{1}^{2}m_{2}^{2}-2m_{1}^{3}m_{2}-2m_{2}^{3}m_{1})\log\lf(2K \ri) \Big\}.
\eea
This shows that the integral diverges in $K$ as $m_{i}^{4}\,\log\lf(
K \ri)$. As shown in Fig.2 the divergence in $K$ is smoothed by
the factor $m_{i}^{4}$. For neutrino masses of order of
$10^{-3}eV$ we have $\rho_{\Lambda}^{mix} = 5.4 \times 10^{-47}GeV^{4}$
for a value of the cut-off of order of the Planck scale $K=10^{19} GeV$.
From Eq.(\ref{darklimit}) one also sees that
$\frac{d \rho_{\Lambda}^{mix}(K)}{d K} \propto \frac{1}{K} \rightarrow 0$
for large $K$.
An interesting question to ask is how the result
 $\rho_{\Lambda}^{mix} \propto m_{i}^{4}\,\log\lf(K \ri)$,
 directly obtained in our approach, is related to the conjecture
 \cite{Sahni:2004ai} that the small value of the cosmological constant
$ \rho_{\Lambda} \propto (10^{-3}eV)^{4}$
is associated with the vacuum in a theory which has a fundamental mass scale
$m \sim 10^{-3}eV$.

\begin{figure}
\centering \resizebox{8.5cm}{!}{\includegraphics{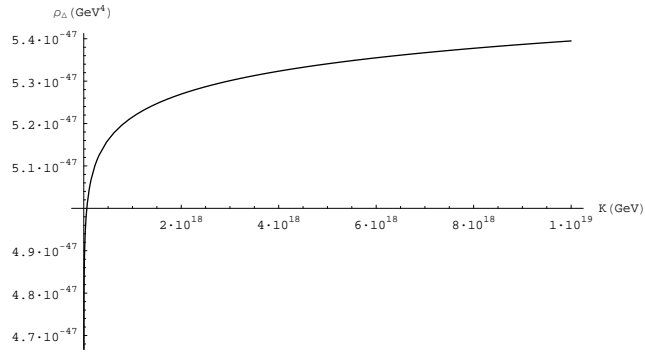}}
\hfill \caption{The neutrino mixing dark energy as a function of
cut-off K.} \label{Fig: 2}
\end{figure}

We observe that, since, at present epoch, the characteristic
oscillation length of the neutrino is much smaller than the
universe curvature radius, the mixing treatment in the flat
space-time, in such an epoch, is a good approximation of that in
FRW space-time. The central result of this paper is: the vacuum
condensate from neutrino mixing can give rise to the {\it
observed} value of the cosmological constant. Exotic components to
dark energy are not necessary in this approach.

\section{Three flavor fermion mixing}

The above result can be easily extended to the three flavor
fermion mixing case.
The Lagrangian density describing three Dirac fields with a mixed
mass term is:
\bea\label{lagemu} {\cal L}(x)\,=\,  {\bar \Psi_f}(x) \lf( i
\not\!\partial - \textsf{M} \ri) \Psi_f(x)\, , \eea
where $\Psi_f^T=(\nu_e,\nu_\mu,\nu_{\tau})$ and $\textsf{M} =
\textsf{M}^\dag$ is the mixed mass matrix.
Among the various possible parameterizations of the mixing matrix
for three fields, we work with CKM matrix of the form:
%
\bea\label{fermix} \Psi_f(x) \, = {\cal U} \, \Psi_m
(x)=\begin{pmatrix}
c_{12}c_{13} & s_{12}c_{13} & s_{13}e^{-i\de} \\
-s_{12}c_{23}-c_{12}s_{23}s_{13}e^{i\de} &
c_{12}c_{23}-s_{12}s_{23}s_{13}e^{i\de} & s_{23}c_{13} \\
s_{12}s_{23}-c_{12}c_{23}s_{13}e^{i\de} &
-c_{12}s_{23}-s_{12}c_{23}s_{13}e^{i\de} & c_{23}c_{13}
\end{pmatrix}\,\Psi_m (x) \, , \eea
with $c_{ij}=\cos\te_{ij}$ and  $s_{ij}=\sin\te_{ij}$, being
$\te_{ij}$ the mixing angle between $\nu_{i},\nu_{j}$ and
$\Psi_m^T=(\nu_1,\nu_2,\nu_3)$.
Using Eq.(\ref{fermix}), we diagonalize the quadratic form of
Eq.(\ref{lagemu}), which then reduces to the Lagrangian for three
Dirac fields, with masses $m_1$, $m_2$ and $m_3$:
\bea\label{lag12} {\cal L}(x)\,=\,  {\bar \Psi_m}(x) \lf( i
\not\!\partial -  \textsf{M}_d\ri) \Psi_m(x)  \, , \eea
where $\textsf{M}_d = diag(m_1,m_2,m_3)$.

The mixing transformation can be written as
$\nu_{\si}^{\al}(x)\equiv G^{-1}_{\bf \te}(t)
\, \nu_{i}^{\al}(x)\, G_{\bf \te}(t), $ where $(\si,i)=(e,1),
(\mu,2), (\tau,3)$, and the generator is now
\bea\label{generator} &&G_{\bf
\te}(t)=G_{23}(t)G_{13}(t)G_{12}(t)\, , \eea
where
\bea\label{generators1} && G_{12}(t)\equiv
\exp\Big[\te_{12}\int
d^{3}x\lf(\nu_{1}^{\dag}(x)\nu_{2}(x)-\nu_{2}^{\dag}(x)\nu_{1}(x)\ri)\Big],
\\ \label{generators2}
&&G_{23}(t)\equiv\exp\Big[\te_{23}\int
d^{3}x\lf(\nu_{2}^{\dag}(x)\nu_{3}(x)-\nu_{3}^{\dag}(x)\nu_{2}(x)\ri)\Big],
\\ \label{generators3}
&&G_{13}(t)\equiv\exp\Big[\te_{13}\int
d^{3}x\lf(\nu_{1}^{\dag}(x)\nu_{3}(x)e^{-i\de}-\nu_{3}^{\dag}(x)
\nu_{1}(x)e^{i\de}\ri)\Big]. \eea

The vacuum for the mass
eigenstates is denoted by $|0\ran_{m}$ and the {\em flavor vacuum} is given by
$|0(t)\ran_{f}\,\equiv\,G_{\te}^{-1}(t)\;|0\ran_{m} \;.$
The flavor annihilation operators in the reference frame ${\bf k}=(0,0,|{\bf
k}|)$ are:
\bea \al_{{\bf k},e}^{r}(t)&=&c_{12}c_{13}\;\al_{{\bf k},1}^{r}(t) +
s_{12}c_{13}\lf(|U^{{\bf k}}_{12}|\;\al_{{\bf k},2}^{r}(t)
+\epsilon^{r} |V^{{\bf k}}_{12}|\;\bt_{-{\bf k},2}^{r\dag}(t)\ri) +
e^{-i\de}\;s_{13}\lf(|U^{{\bf k}}_{13}|\;\al_{{\bf k},3}^{r}(t)
+\epsilon^{r} |V^{{\bf k}}_{13}|\;\bt_{-{\bf k},3}^{r\dag}(t)\ri)\;,
\\[2mm]\non
\al_{{\bf k},\mu}^{r}(t)&=&\lf(c_{12}c_{23}- e^{i\de}
\;s_{12}s_{23}s_{13}\ri)\;\al_{{\bf k},2}^{r}(t) -
\lf(s_{12}c_{23}+e^{i\de}\;c_{12}s_{23}s_{13}\ri) \lf(|U^{{\bf
k}}_{12}|\;\al_{{\bf k},1}^{r}(t) -\epsilon^{r} |V^{{\bf
k}}_{12}|\;\bt_{-{\bf k},1}^{r\dag}(t)\ri)
\\
&&+\;s_{23}c_{13}\lf(|U^{{\bf k}}_{23}|\;\al_{{\bf k},3}^{r}(t) +
\epsilon^{r} |V^{{\bf k}}_{23}|\;\bt_{-{\bf k},3}^{r\dag}(t)\ri)\;,
\\[2mm]\non
\al_{{\bf k},\tau}^{r}(t)&=&c_{23}c_{13}\;\al_{{\bf k},3}^{r}(t) -
\lf(c_{12}s_{23}+e^{i\de}\;s_{12}c_{23}s_{13}\ri) \lf(|U^{{\bf
k}}_{23}|\;\al_{{\bf k},2}^{r}(t) -\epsilon^{r} |V^{{\bf
k}}_{23}|\;\bt_{-{\bf k},2}^{r\dag}(t)\ri)
\\
&&+\;\lf(s_{12}s_{23}- e^{i\de}\;c_{12}c_{23}s_{13}\ri)
\lf(|U^{{\bf k}}_{13}|\;\al_{{\bf k},1}^{r}(t) -\epsilon^{r} |V^{{\bf
k}}_{13}|\;\bt_{-{\bf k},1}^{r\dag}(t)\ri)\;, \eea

\bea
\bt^{r}_{-{\bf k},e}(t)&=&c_{12}c_{13}\;\bt_{-{\bf k},1}^{r}(t)
+ s_{12}c_{13}\lf(|U^{{\bf k}}_{12}|\;\bt_{-{\bf k},2}^{r}(t)
-\epsilon^{r}|V^{{\bf k}}_{12}|\;\al_{{\bf k},2}^{r\dag}(t)\ri)
+e^{i\de}\; s_{13}\lf(|U^{{\bf k}}_{13}|\;\bt_{-{\bf k},3}^{r}(t)
-\epsilon^{r} |V_{13}^{{\bf k}}|\;\al_{{\bf k},3}^{r\dag}(t)\ri)\;,
\\[2mm] \non
\bt^{r}_{-{\bf k},\mu}(t)&=&\lf(c_{12}c_{23}- e^{-i\de}\;
s_{12}s_{23}s_{13}\ri)\;\bt_{-{\bf k},2}^{r}(t)-
\lf(s_{12}c_{23}+e^{-i\de}\;c_{12}s_{23}s_{13}\ri) \lf(|U^{{\bf
k}}_{12}|\;\bt_{-{\bf k},1}^{r}(t) +\epsilon^{r}\; |V^{{\bf
k}}_{12}|\;\al_{{\bf k},1}^{r\dag}(t)\ri) +
\\
&&+\; s_{23}c_{13}\lf(|U^{{\bf k}}_{23}|\;\bt_{-{\bf k},3}^{r}(t) -
\epsilon^{r}\; |V^{{\bf k}}_{23}|\;\al_{{\bf k},3}^{r\dag}(t)\ri)\;,
\\[2mm] \non
\bt^{r}_{-{\bf k},\tau}(t)&=&c_{23}c_{13}\;\bt_{-{\bf k},3}^{r} -
\lf(c_{12}s_{23}+e^{-i\de}\;s_{12}c_{23}s_{13}\ri) \lf(|U^{{\bf
k}}_{23}|\;\bt_{-{\bf k},2}^{r}(t) + \epsilon^{r} |V^{{\bf
k}}_{23}|\;\al_{{\bf k},2}^{r\dag}(t)\ri)
\\
&&+\;\lf(s_{12}s_{23}- e^{-i\de}\;c_{12}c_{23}s_{13}\ri)
\lf(|U^{{\bf k}}_{13}|\;\bt_{-{\bf k},1}^{r}(t) + \epsilon^{r}
|V^{{\bf k}}_{13}|\;\al_{{\bf k},1}^{r\dag}(t)\ri)\;.
\eea

These operators satisfy canonical (anti)commutation relations at
equal times. $U^{{\bf k}}_{ij}$
and $V^{{\bf k}}_{ij}$ are Bogoliubov coefficients defined as:

\bea &&|U^{{\bf
k}}_{ij}|=\lf(\frac{\om_{k,i}+m_{i}}{2\om_{k,i}}\ri)
^{\frac{1}{2}}
\lf(\frac{\om_{k,j}+m_{j}}{2\om_{k,j}}\ri)^{\frac{1}{2}}
\lf(1+\frac{|{\bf k}|^{2}}{(\om_{k,i}+m_{i})
(\om_{k,j}+m_{j})}\ri)
\\
&&|V^{{\bf k}}_{ij}|=\lf(\frac{\om_{k,i}+m_{i}}{2\om_{k,i}}\ri)
^{\frac{1}{2}}
\lf(\frac{\om_{k,j}+m_{j}}{2\om_{k,j}}\ri)^{\frac{1}{2}}
\lf(\frac{|{\bf k}|}{(\om_{k,j}+m_{j})}-\frac{|{\bf
k}|}{(\om_{k,i}+m_{i})}\ri)
 \eea
  \bea
|U^{{\bf k}}_{ij}|^{2}+|V^{{\bf k}}_{ij}|^{2}=1 \eea where
$i,j=1,2,3$ and $j>i$.
The condensation densities are different for particles of different masses:
\bea {\cal N}^{\bf k}_1\, = \,_{f}\langle0(t)|N^{{\bf
k},r}_{\al_{1}} |0(t)\ran_{f}&=& \,_{f}\langle0(t)|N^{{\bf
k},r}_{\bt_{1}}|0(t)\ran_{f}= s^{2}_{12}c^{2}_{13}\,|V^{{\bf
k}}_{12}|^{2}+ s^{2}_{13}\,|V^{{\bf k}}_{13}|^{2}\,,
\\
{\cal N}^{\bf k}_2\, = \,_{f}\langle0(t)|N^{{\bf
k},r}_{\al_{2}}|0(t)\ran_{f}&=& \,_{f}\langle0(t)|N^{{\bf
k},r}_{\bt_{2}}|0(t)\ran_{f}
=\lf|-s_{12}c_{23}+e^{i\de}\,c_{12}s_{23}s_{13}\ri|^{2} \,|V^{{\bf
k}}_{12}|^{2}+ s^{2}_{23}c^{2}_{13}\;|V^{{\bf k}}_{23}|^{2}\,,
\\
{\cal N}^{\bf k}_3\, = \,_{f}\langle0(t)|N^{{\bf
k},r}_{\al_{3}}|0(t)\ran_{f}&=& \,_{f}\langle0(t)|N^{{\bf
k},r}_{\bt_{3}}|0(t)\ran_{f}=
\lf|-c_{12}s_{23}+e^{i\de}\,s_{12}c_{23}s_{13}\ri|^{2} |V^{{\bf
k}}_{23}|^{2} + \lf|s_{12}s_{23}+
e^{i\de}\,c_{12}c_{23}s_{13}\ri|^{2} |V^{{\bf k}}_{13}|^{2}\,.
\eea

In this case, at present epoch, the contribution given to the dark energy
by the neutrino mixing is
 \bea \label{cost1}\non
\rho_{\Lambda}^{mix} & = & \frac{2}{\pi}  \int_{0}^{K} dk \, k^{2}
\Big[\frac{m_{1}^{2}}{\omega_{k,1}}\lf(s^{2}_{12}c^{2}_{13}\,|V^{{\bf
k}}_{12}|^{2}+ s^{2}_{13}\,|V^{{\bf k}}_{13}|^{2}\ri) +
\frac{m_{2}^{2}}{\omega_{k,2}}\lf(\lf|-s_{12}c_{23}+e^{i\de}\,c_{12}s_{23}s_{13}\ri|^{2}
\,|V^{{\bf k}}_{12}|^{2}+ s^{2}_{23}c^{2}_{13}\;|V^{{\bf
k}}_{23}|^{2}\ri)
\\ & + & \frac{m_{3}^{2}}{\omega_{k,3}}\lf(\lf|-c_{12}s_{23}+e^{i\de}\,s_{12}c_{23}s_{13}\ri|^{2} |V^{{\bf
k}}_{23}|^{2} + \lf|s_{12}s_{23}+
e^{i\de}\,c_{12}c_{23}s_{13}\ri|^{2} |V^{{\bf k}}_{13}|^{2} \ri)
\Big] \,,
 \eea
which can be written as
\bea \label{cost2}\non \rho_{\Lambda}^{mix}
& = & \frac{2}{\pi}  \int_{0}^{K} dk \, k^{2}
\Big\{\frac{m_{1}^{2}}{\omega_{k,1}}\lf(s^{2}_{12}c^{2}_{13}\,|V^{{\bf
k}}_{12}|^{2}+ s^{2}_{13}\,|V^{{\bf k}}_{13}|^{2}\ri) +
\frac{m_{2}^{2}}{\omega_{k,2}}\lf[\lf(s^{2}_{12}c^{2}_{23} +
 c^{2}_{12}s^{2}_{23}s^{2}_{13}\ri) \,|V^{{\bf
k}}_{12}|^{2}+ s^{2}_{23}c^{2}_{13}\;|V^{{\bf k}}_{23}|^{2}\ri]
\\\non & + & \frac{m_{3}^{2}}{\omega_{k,3}}\lf[\lf(c^{2}_{12}s^{2}_{23} + s^{2}_{12}c^{2}_{23}s^{2}_{13}\ri)
 |V^{{\bf k}}_{23}|^{2} + \lf(s^{2}_{12}s^{2}_{23}+
c^{2}_{12}c^{2}_{23}s^{2}_{13}\ri) |V^{{\bf k}}_{13}|^{2} \ri]
\Big\}
\\ & - & \frac{4}{\pi} s_{12}c_{23}c_{12}s_{23}s_{13}c_{\delta} \int_{0}^{K}
dk \, k^{2} \Big\{ \frac{m_{2}^{2}}{\omega_{k,2}}  \,|V^{{\bf
k}}_{12}|^{2}
 +  \frac{m_{3}^{2}}{\omega_{k,3}}\lf[|V^{{\bf k}}_{23}|^{2} - |V^{{\bf k}}_{13}|^{2} \ri] \Big\}
 \,,
 \eea
where $c_{\delta}= \cos \delta$.
We note that $\rho_{\Lambda}^{mix}$ is also depending on the $CP$
violating phase $\delta$. Like in the case of two flavor neutrino
mixing, the integral diverges in $K$ as $m_{i}^{4}\,\log\lf( K
\ri)$. A value of $ \rho_{\Lambda}^{mix}$, compatible with the
upper bound on the cosmological dark energy, is obtained for
neutrino masses of order of $10^{-3}eV$ so the result is
essentially the same of the two flavor case.

\section{Conclusions and discussion}

The vacuum condensate generated by neutrino mixing can be interpreted
as an evolving dark energy that, at present epoch, behaves as the cosmological constant,
giving rise to its observed value.
The result is naturally achieved even when a cut-off $K$
of the order of Planck scale is considered. It is easily recovered
also for three flavor fermion mixing. Such
a result links together dark energy with the neutrino masses.
Introducing auxiliary fields or mechanisms is not required in our approach.

A short summary of the observational status of art  can aid to
clarify the frame for our considerations and results. An increasing
bulk of data have been accumulated in the last few years.
They have paved the way to the emergence of a new standard
cosmological model usually referred to as the {\it concordance
model}. The Hubble diagram of Type Ia Supernovae (SNeIa), measured
by both the Supernova Cosmology Project \cite{SCP} and the
High\,-\,z Team \cite{HZT} up to redshift $z \sim 1$, was the
first evidence  that the universe is undergoing a phase of
accelerated expansion. On the other hand, balloon born
experiments, such as BOOMERanG \cite{Boomerang} and MAXIMA
\cite{Maxima}, determined the location of the first and second
peak in the anisotropy spectrum of  cosmic microwave background
radiation (CMBR)  pointing out that the geometry of the universe
is spatially flat. If combined with constraints coming from galaxy
clusters on the matter density parameter $\Omega_M$, these data
indicate that the universe is dominated by a non-clustered fluid
with negative pressure, generically dubbed {\it dark energy},
which is able to drive the accelerated expansion. This picture has
been further strengthened by the more precise measurements of the
CMBR spectrum, due to the WMAP experiment \cite{WMAP}, and by the
extension of the SNeIa Hubble diagram to redshifts higher than 1
\cite{Riess04}. Several models trying to
explain this phenomenon have been presented; the simplest
explanation is claiming for the well known cosmological constant
$\Lambda$ \cite{LCDMrev}. Although the best fit to most of the
available astrophysical data \cite{WMAP}, the $\Lambda$CDM model
fails in explaining why the inferred value of $\Lambda$ is so tiny
(120 orders of magnitude lower) compared to the typical vacuum
energy values predicted by particle physics and why its energy
density is today comparable to the matter density (the so called
{\it coincidence problem}). As a tentative solution, many authors
have replaced the cosmological constant with a scalar field
rolling down its potential and giving rise to  models  referred to
as {\it quintessence} \cite{QuintRev}. Even if successful in
fitting the data, the quintessence approach to dark energy is
still plagued by the coincidence problem since the dark energy and
matter densities evolve differently and reach comparable values
for a very limited portion of the universe evolution  coinciding
at present era. In this case, the coincidence problem is replaced
with a fine-tuning problem. Moreover, it is not clear where this
scalar field originates from, thus leaving a great uncertainty on
the choice of the scalar field potential.

The subtle and elusive
nature of  dark energy has led to  look for
completely different scenarios able to give a quintessential
behavior without the need of exotic components. To this aim, we
observe that the acceleration of the universe only claims
for a negative pressure dominant component, but does not tell
anything about the nature and the number of cosmic fluids filling
the universe \cite{Capozziello:2006dj}. This consideration suggests that it could be
possible to explain the accelerated expansion by introducing a
single cosmic fluid with an equation of state causing it to act
like dark matter at high densities (giving rise to clustered
structures)  and dark energy at low densities (then giving rise to
accelerated behavior of cosmic fluid).  An attractive feature of
these models, usually referred to as {\it Unified Dark Energy}
(UDE) or {\it Unified Dark Matter} (UDM) models, is that such an
approach naturally solves, at least phenomenologically, the
coincidence problem. Some interesting examples are the generalized
Chaplygin gas \cite{Chaplygin}, the tachyon field \cite{tachyon}
and the condensate cosmology \cite{Bassett}. A different class of
UDE models has been proposed \cite{Hobbit} where a single fluid is
considered whose energy density scales with the redshift in such a
way that the radiation dominated era, the matter dominated era and
the accelerating phase can be naturally achieved.
 Actually, there is still a
different way to face the problem of cosmic acceleration.
 It is possible that the
observed acceleration is not the manifestation of another
ingredient in the cosmic pie, but rather the first signal of a
breakdown of our understanding of the laws of gravitation \cite{CCT,garattini} .
Examples of models comprising only the standard matter are provided by
the Cardassian expansion
\cite{Cardassian}, the DGP gravity \cite{DGP},
higher order gravity actions \cite{curvature}, non\,-\,vanishing torsion field
 \cite{torsion}, higher-order curvature invariants included in the gravity
Lagrangian \cite{curvfit}, etc..

This abundance of
models is from one hand the signal of the fact that we have a
limited number of cosmological tests to discriminate among rival
theories, and from the other hand, that a urgent degeneracy
problem has to be faced. The evidences of neutrino
oscillations \cite{SNO,K2K} and the fact that the vacuum condensate
 originated by  neutrino mixing
provides contributions to the dark energy compatible with today
expected value, as shown in the present paper, could contribute towards a solution of
such a problem from both  experimental and theoretical viewpoints.

\section*{Acknowledgements}

One of the authors (A.C.) acknowledges the Department of Physics and Astronomy,
University of Leeds for partial financial support.
Support from INFN and MURST is also acknowledged.

\bibliography{apssamp}

\end{document}